# *Bivalent nature of Fe and W ions in the antiferromagnetic $Sr_2FeWO_6$ material produced by the combustion technique*

J. I. Villa Hernández[a], L. Suescun[b], D. R. Sanchez[c], J. Grassi[b], N. Di Benedetto[b], S. C. Tidrow[d], D. A. Landínez Téllez[a], J. Roa-Rojas[a,1]

a. Grupo de Física de Nuevos Materiales, Departamento de Física, Universidad Nacional de Colombia, Bogotá DC, Colombia
b. Cryssmat-Lab/Cátedra de Física/DETEMA, Facultad de Química, Universidad de la República, Montevideo 11800, Uruguay
c. Instituto de Física, Universidade Federal Fluminense, 24210-346 Niterói, RJ, Brazil
d. Inamori School of Engineering, Alfred University, Alfred, New York, USA.

**Abstract**

The synthesis of the double perovskite $Sr_2FeWO_6$ through the direct combustion technique is reported. A complete characterization of the crystalline structure is performed from Synchrotron X-ray measurements and subsequent Rietveld analysis. The results evidence that this material crystallizes in a monoclinic structure, $P2_1/n$ space group, with strong distortions in the $Fe-O_6$ and $W-O_6$ octahedra, which is consistent with Glazer's notation $a^-b^+a^-$. A superficial morphological study through scanning electron microscopy images reveals that this synthesis technique allows obtaining nanometric grains with *160±5 nm* mean size. The semiquantitative compositional analysis through spectroscopy of X-ray energy dispersion by electrons shows that the produced material does not contain other chemical elements besides those expected form the stoichiometric formula. Results of analyzes of spectroscopy of photoelectrons emitted by X-rays suggest that there is 66.9% of the material with oxidation states $Fe^{2+}$ and $W^{6+}$ and 33.1% with oxidation states $Fe^{3+}$ and $W^{5+}$. This bivalence of Fe and W cations is corroborated by means the Mössbauer spectroscopy technique. Finally, the magnetic response in AC susceptibility curves allows us to affirm that at low temperatures the material adopts an antiferromagnetic behavior with $T_N=32\ K$, with a tendency to the ferromagnetic state above *T=150 K* mostly due to the presence of $Fe^{3+}$ ions in the high spin configuration $3d^5$, $S = 5/2$.

---

[1] Corresponding author: jroar@unal.edu.co

**Research Areas:** Antiferromagnetism; Crystal structure; Crystal symmetry; Ferromagnetism; Structural properties.

I. **Introduction.**

During the last four decades the interest in perovskite-type materials has been so great and growing, that it is difficult to cite the most important works because the relevance of each investigation has depended on the specific properties studied [1]. The versatility of this family of ceramic materials lies in the possibility of including many elements of the periodic table within its ideal chemical formula $ABX_3$, where A is usually an alkaline earth or rare earth, B a transition metal or a lanthanide and X usually is oxygen [2]. Within the options of modification of the perovskites, a very important one is the double structure given by the formula $A_2BB'O_6$, which allows the introduction of an additional cation, facilitating the occurrence of dramatic changes in the physical properties of the material, which are from insulators, semiconductors and conductors (from the point of view of electrical transport) to paramagnetic, ferromagnetic, ferrimagnetic and antiferromagnetic (from the point of view of the magnetic ordering), allowing the obtaining of novel ferroelectric, multiferroic, colossal magnetoresistive and half- metallic materials, among others [3-6]. An interesting perovskite that has been studied due to its magnetoresistive response and because it evidences a metal-insulator transition accompanied by a transition from charge-ordered to orbital-ordered is $SrFeO_3$ [7-8]. This simple perovskite was reported to crystallize in a cubic structure, independent of temperature, with metallic electrical response [9-10] and the theoretical calculations of electronic band structure do not reveal the occurrence of the Jahn Teller effect [11]. On the other hand, the magnetoresistive response can become negative or positive, depending on the deficiency of oxygen, but does not seem to involve mechanisms of ferromagnetic ordering, which gives it an exotic and interesting character [12]. These unique characteristics of the $SrFeO_3$ material justify the inclusion of transition metals in the B sites of their structure, forming double perovskites of the $Sr_2FeB'O_6$ type, which has been done for more than five decades with B'=W [13]. Meanwhile, the author reported the cubic structure of the $Sr_2FeWO_6$ without considering the formation of a salt-rock superstructure that is characteristic of this family of double perovskites with cationic ordering, which would be constituted by 8 cubes containing the Sr cation, each one of them octahedrally coordinated with 6 oxygen anions, and the Fe and W cations intercalary located along the three crystallographic axes [14]. The large magnetoresistive nature of $Sr_2FeMoO_6$ [15], as well as its half-metallic character [16], accelerated the studies of $Sr_2FeWO_6$ due to the relative similarity between Mo and W, since they belong to the same group of the periodic table. A short report suggests that this material behaves as a semiconductor and antiferromagnetic at temperatures below 37 K, with the iron ground state being high spin $Fe^{2+}$ [17]. However, although the authors describe characteristic cell parameters corresponding to a tetragonal structure, they do not specifically report the space group. In a slightly more recent report, the $Sr_2FeWO_6$ is indexed as belonging to the $P2_1/n$ space group, which corresponds to a monoclinic structure with cell parameters $a \neq b \neq c$ [18]. The authors claim that antiferromagnetic type coupling takes place in this material at low temperatures due to the presence of unpaired electrons in the crystallographic iron sites. Some calculations of the electronic structure suggest that the insulating character at high temperatures is due to the fact that the hybridization between 2p-O states and 5d-W states is strong enough to push the 5d band higher in energy, opening the band interval to inhibit the transfer of electrons [19-20]. Studies of the valence state in $Sr_2FeWO_6$ by means of X-Ray absorption near-edge spectra suggest that it mainly has $Fe^{2+}$ and $W^{6+}$, with small traces of $W^{5+}$ [21]. In the meantime, because the interpretation of magnetic behavior and the insulating response of this material is an open issue, some relatively recent works in which the content of Mo by W in the double perovskite $Sr_2FeMo_{1-x}W_xO_6$ is partially replaced have been reported [22-23]. Some of the reports made for both $Sr_2FeWO_6$ and $Sr_2FeMo_{1-x}W_xO_6$ have to do with samples produced by solid reaction [24] and sol-gel [22]. It is well known that some physical properties of perovskite-type materials depend on the synthesis technique, since surface morphology substantially influences their electrical and magnetic responses [25]. With the aim of contributing to this interesting discussion, in this work the synthesis through the combustion technique, as well as the structural, morphological, resistive and magnetic characterization of the $Sr_2FeWO_6$ material are reported.

## II.  Experimental setup

Samples of $Sr_2FeWO_6$ were produced by the combustion technique, from the precursor Strontium nitrate $Sr(NO_3)_2$, Iron (III) nitrate $Fe(NO_3)_3(H_2O)_9$, and Tungstic acid $H_2WO_4$ (all nitrates with purities greater than 99%), forming a solution, the first three in distilled water and the last in ethanol. EDTA (Ethylenediaminetetraacetic acid) was used as a chelating agent and ammonium nitrate ($NH_4NH_3$) as fuel, such that the ratio metal ions: EDTA: $NH_4NH_3$ was 4:4:10. Subsequently, the EDTA + $NH_4NH_3$ solution was added and its temperature was increased on a heating plate (up to *T≈120 °C*) with continuous magnetic stirring, verifying that its pH value was greater than 10, until obtaining a gel. Afterwards, the gel temperature was raised (up to ≈ *300 °C*) to promote the self-combustion process. The ashes resulting from the combustion process were subjected to a macerate in an agate mortar in order to homogenize the obtained material. It should be noted that it was not necessary to subject the material to any additional heat treatment or to use environments rich in inert gases to obtain the material in the majority crystalline phase. The processes involved in the synthesis of the material by conventional methods include factors that increase energy expenditure and limit synthesis processes to assemblies under reducing atmospheres. For this reason, the assisted gel combustion method appears as a viable and very helpful alternative when seeking to obtain materials with a perovskite type structure that need special synthesis conditions. [20-21]. The crystalline structure of the materials was studied through the X-ray diffraction technique (XRD), from diffraction patterns that were taken at the Brazilian Synchrotron Light Source (LNLS), Campinas, Brazil, which has a radiation wavelength of *1.2372 Å*. The data were collected in the range *15° ≤ 2θ ≤ 120°*, by using a linear collector with a step of *0.025°* and an integration of *0.005* in *t=2 s* exposition time. The analyses of the diffraction patterns were carried out using the GSAS-II software [26-27]. The morphological characterization of the materials was evaluated through scanning electron microscopy (SEM) images, through a FEI Quanta 200 microscope, at magnifications of 10 kx, 20 kx and 36 kx. The semiquantitative compositional analysis of the samples under study was carried out using the Energy-dispersive X-ray spectroscopy (EDX) technique. The measurements of Mössbauer spectroscopy were performed in the transmission geometry, using a $^{57}$Fe source. The spectra were adjusted using the NORMOS software. For the measurements at low temperature an OXFORD cryostat was used, with the radioactive source following a sinusoidal movement. During measurement, both the radioactive source and the sample remain at the same temperature as the sample. Analyses of spectroscopy of photoelectrons emitted by X-rays (XPS) were performed, using a Quantera SXM Scanning X-Ray microscope with an Al Kα X-ray monochromatic source (*1486.7 eV*), with a beam size of *100 μm* and a Hemispheric Electron Energy Analyzer (HSA). The pressure of the base of the chamber was $3x10^9$ *Torr*. The energy spectra were taken in the range from *0* to *1200 eV*, with a step of *0.5 eV* and were analyzed by the CASAXPS software [28]. The C 1s peak (*179.15 eV*) was used as an internal standard to correct the maximum changes due to the accumulation of surface charge in samples. The magnetic properties were studied by means of AC susceptibility measurements through a Quantum Design MPMS (Magnetic Properties Measurement System), on the application of *H=10 Oe* magnetic field at a frequency of *1000 Hz*, in the temperature range *6 K ≤ T ≤ 300 K*.

**III. Results and discussion**

*A. Structural characterization*

The XRD pattern presented in figure 1 corresponds to the refined experimental data using the GSAS-II code, where the continuous curve corresponds to the pattern calculated by means of the GSAS code and the symbols represent the data obtained experimentally. In the Rietveld analysis process of the experimental pattern, the scale factors, atomic positions $(x,y,z)$ and lattice parameters $(a,b,c)$ of each of the crystalline phases present in the material were refined, as well as peak width parameters $(u,v,w)$, profile functions, instrumental parameters related to the XRD equipment and the function that models the background.

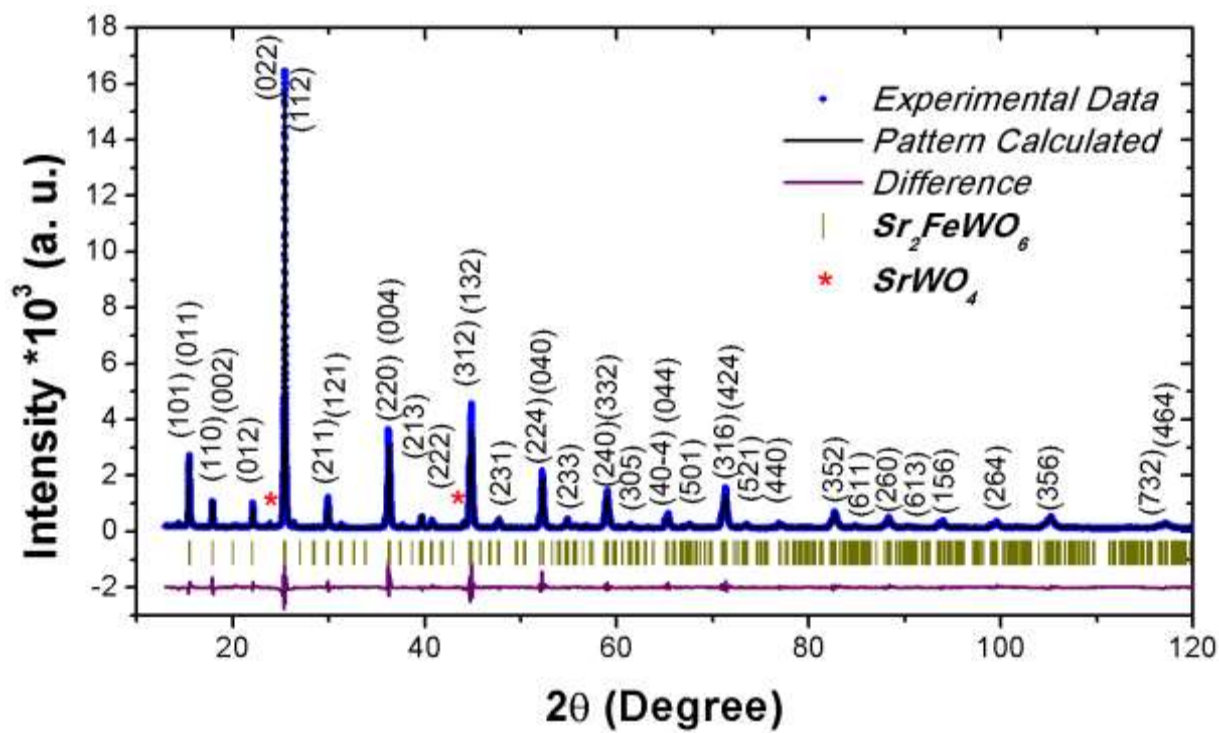

Fig. 1. Refined diffractogram for the $Sr_2FeWO_6$ double perovskite produced through the combustion technique.

In the lower part of the diffractogram, the continuous line exemplifies the difference between the simulated data and the experimental values, while the short vertical lines show the angular positions of the Bragg peaks corresponding to the observed structural phase of the $Sr_2FeWO_6$ compound. The apparent great difference between experimental and simulated patterns is due to the tensions that occur between the particles and that are a consequence of the synthesis method. The two asterisks marked in the pattern correspond to a small phase of oxide material that did not enter the structure of the majority phase.

The Rietveld refinement of the experimental data of XRD suggests that at room temperature the material evidences a majority crystallographic phase of the $Sr_2FeWO_6$ (96.5%) with a centrosymmetric primitive monoclinic structure of the perovskite type, belonging to the spatial group *$P2_1/n$ (# 14)*. Another very small phase (3.5%), corresponding to the $SrWO_4$ compound, was identified as a result of the excellent resolution in the performed measurements and the quality of the Rietveld analysis. This ternary oxide was equally refined, finding that it crystallized in a Scheelite tetragonal structure identified by the space group *$I4_1/a$*. The presence of diffraction planes (011), (211) and (231), among others, in the diffractogram of fig. 1 for the majority phase, constitutes a fingerprint of the conformation of an ordered arrangement of the Fe and W cations, which characterizes the superstructure of the complex perovskites $A_2BB'O_6$ [29]. The confirmation of cationic ordering given by the superstructure allows establishing the occurrence of an alternate distribution of the Fe and W cations in sites B and B' of the double perovskite, each of them coordinated with 6 oxygens, forming octahedral substructures. Sr cation is located at site A of the complex structure. This arrangement of the ions in the unit cell is presented in figure 2.

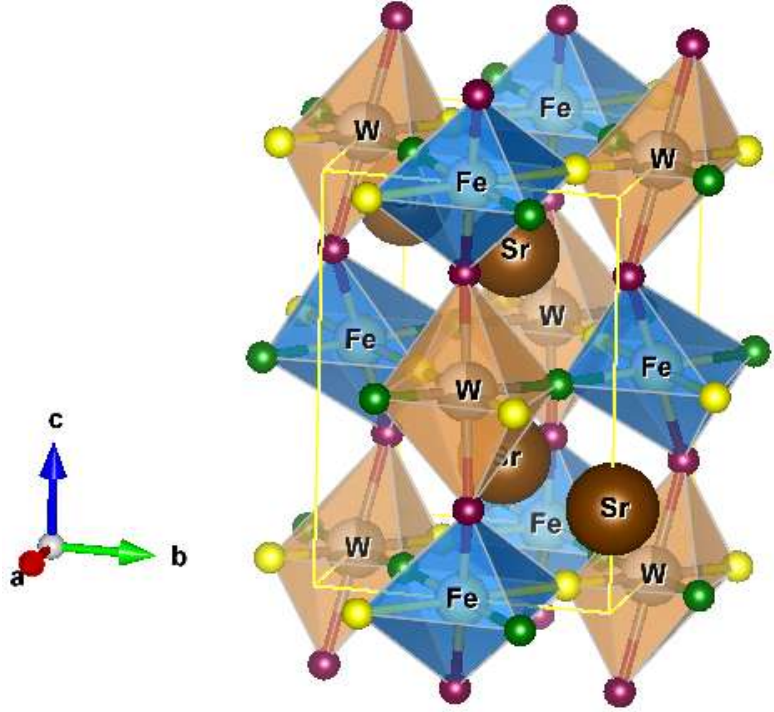

Fig. 2. Crystal structure of the $Sr_2FeWO_6$ complex perovskite in the $P2_1/n$ (#14) space group.

The box outlined in the figure describes the unit cell, with primitive cell parameter $a_p$, from which the lattice parameters are expected $a \approx \sqrt{2}a_p, b \approx \sqrt{2}a_p$ and $c \approx 2a_p$. Another important aspect in the crystal structure of the perovskite complex $Sr_2FeWO_6$ belonging to the *P2₁/n* space group is the distorted character of the structure with respect to the ideal cubic perovskite. This distortion has to do with the so-called tolerance factor *τ<1*, for which the structure has less symmetry. The characteristic τ can be calculated by the expression

$$\tau = \frac{R_A + R_O}{\sqrt{2}\left[\frac{(R_B + R_{B\prime})}{2} + R_O\right]}, \quad (1)$$

where $R_A$, $R_O$, $R_B$ and $R_{B'}$ are the ionic radii of the A, O, B and B' for the generic formula $A_2BB'O_6$. Thus, in the particular case of the $Sr_2FeWO_6$ material, $R_A=R_{Sr}$, $R_B=R_{Fe}$, and $R_{B'}=R_W$. The value of the tolerance factor was *τ=0.961(7)*, which is relatively far from the expected value for an ideal cubic perovskite. However, it is necessary to consider that the deviations with respect to the ideal cubic structure, from the point of view of the tolerance factor, have recently been discussed and redefined. [30]. Meanwhile, it is expected for the P2$_1$/n space group to show octahedral distortions, since in Glazer's notation this one corresponds to a system with tilting given by $a^-b^+a^-$, representing a non-symmetric spatial group. Another important concept in the structural study of new materials is position. In the present analysis the Wyckoff positions obtained for the structure of the Sr$_2$FeWO$_6$ double perovskite are shown in table I.

Table I. Wyckoff positions and tolerance factor obtained from the Rietveld refinement for the Sr$_2$FeWO$_6$ perovskite compound.

| Atom | Wyckoff Position | x | y | z |
|---|---|---|---|---|
| Sr$^{2+}$ | 4e | -0.0242(5) | -0.0045(1) | 0.7613(5) |
| Fe$^{2+}$ | 2c | 0.5000(0) | 0.0000(0) | 0.5000(0) |
| W$^{6+}$ | 2d | 0.5000(0) | 0.0000(0) | 0.0000(0) |
| O$^{2-}$(1) | 4e | 0.3017(0) | 0.2265(0) | -0.0246(0) |
| O$^{2-}$(2) | 4e | 0.2857(0) | 0.7676(0) | -0.0267(0) |
| O$^{2-}$(3) | 4e | 0.2855(0) | 0.2135(0) | 0.7675(0) |

In the notation used here, *c*, *d* and *e* are the Wyckoff letters that determine all points x for which the site symmetry groups are conjugated subgroups of P2$_1$/n [31]. These letters only represent a coding system for the Wyckoff positions that start with one in the lower position and continue in alphabetical order. The numbers of equivalent points per unit cell, which are next to Wyckoff's letters, is known as the multiplicity of Wyckoff's position [32]. The various values at the Wyckoff positions of the oxygen ions are an obvious signal of octahedral distortions, which can also be stated by observing that the Fe and W cations occupy the different positions *2c* and *2d*, respectively. Figure 3 illustrates the octahedral distortions through a view in the [001] direction (*ab* plane) of the unit cell of the Sr$_2$FeWO$_6$ in the *P2$_1$/n* monoclinic structure. On the other hand, the occurrence of peaks with diffraction conditions *h0l, h+l=2n, 0k0* with *k=2n* including *hkl, 0kl, hk0, h00* and *00l* without any condition, allow to establish that the spatial group chosen for the indexation of the material is right. [33-34]. Likewise, figure 3a allows observing the type of octahedral distortion implicit in Glazer's notation $a^-b^+a^-$. The size differences between the Fe-O$_6$ and W-O$_6$ octahedra can also be clearly seen in figures 3a and 3b, which are related to the different values in ionic radii of these cations as well as to the differences between the hybridizations of the cation Fe with the anion O and the cation W with the anion O. These differences take place because of the different oxidation numbers of the Fe and W cations, as we will see later, will analyze these characteristics through Mössbauer and XPS spectroscopies.

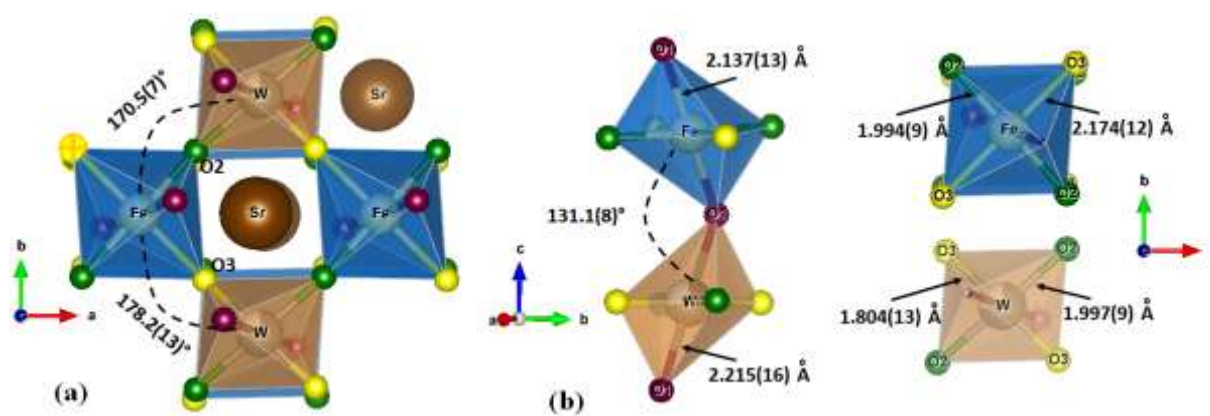

Fig. 3. (a) Tilting of the FeO$_6$ and WO$_6$ octahedra around the [001] direction of the unit cell for the Sr$_2$FeWO$_6$ structure, and (b) inter-atomic distances of the Fe-O and W-O bonds in the octahedra.

In the same way, in both figures 3a and 3b, it is possible to identify significant differences between the tilting angles of the Fe-O$_6$ and W-O$_6$ octahedra, and between the distances of the Fe-O and W-O bonds throughout the three octahedral axes. In Table II, the lengths of the interatomic bonds, the angles between the interatomic bonds and other relevant structural parameters that were obtained by Rietveld refinement of the experimental X-ray diffraction data are summarized for the Sr$_2$FeWO$_6$ complex perovskite. Among the results of table I, it is interesting to note that the lattice parameters *a*, *b* and *c*, as well as the *β* monoclinic angle, are in agreement up to 99.9% with another report for the Sr$_2$FeWO$_6$ material produced through the ceramic technique and analyzed from neutron powder diffraction [18]. In summary, both the Wyckoff positions presented in table I and the interatomic distances and inter-atomic bond angles specified in table II suggest the occurrence of strong octahedral distortions with substantial differences between the sizes and orientations of the Fe-O$_6$ and W-O$_6$ octahedra, which have a significant influence on the physical properties of the material.

Table II. Structural parameters of the Sr$_2$FeWO$_6$ double perovskite obtained by the Rietveld refinements of the X-ray diffraction data for the P2$_1$/n space group.

| Main unit cell parameters obtained from Rietveld refinements of the X-Ray diffraction (XRD) patterns | | | | | |
|---|---|---|---|---|---|
| $a=5.6429(7)$ Å | $b=5.6097(6)$ Å | $c=7.9330(3)$ Å | $\beta=90.061(5)°$ | Vol=251.125 Å$^3$ | $\tau=0.961(7)$ |

| Statistics of the Rietveld refinement | | | | | |
|---|---|---|---|---|---|
| | $X^2=4.08$ | $R_F=2.28\%$ | $R_{wp}=9.31\%$ | $R\_B_{kg}=8.84\%$ | $R=6.77\%$ |

| Main inter-atomic bond lengths | | | | | |
|---|---|---|---|---|---|
| Cation | Anion | Multiplicity | Distance (Å) | Main bond angles (°) | |
| Fe | O1 | x2 | 2.30(4) | Fe-O1-W | 167(3) |
| Fe | O2 | x2 | 2.21(4) | Fe-O2-W | 167(4) |
| Fe | O3 | x2 | 2.71(4) | Fe-O3-W | 98.5(1) |
| W | O1 | x2 | 1.70(4) | O1-Fe-O1 | 180 |
| W | O2 | x2 | 1.80(4) | O2-Fe-O2 | 180 |
| W | O3 | x2 | 2.51(4) | O3-Fe-O3 | 180 |

*B. Surface microstructure and composition*

The shape and distribution of granular agglomerates and particle size are a relevant issue in the study of the microscopic structure and surface morphology in perovskite-type materials, since the particle size and the homogeneity of the microstructural distribution affect the properties of the material. It is known that the microstructure of the sintered ceramic type samples dramatically influences the corrosion resistance and the thermal, magnetic, electrical and mechanical properties [35]. Particularly, in materials of the perovskite kind it has been reported that the dielectric and ferroelectric properties are strongly dependent on the crystal structure and the particle sizes [36]. The representative surface morphology of the Sr$_2$FeWO$_6$ samples is showed in the micrographs of figure 4. Figure 4a, for a magnification of 10 kx, reveals a granular distribution growing in the form of large spongy sheets characterized by a high porosity. When the magnification is doubled at 20 kx (Figure 4b), it is observed that the granular particles are of much smaller sizes than the scale of the image and extend throughout the material, coupling with each other in a random manner.

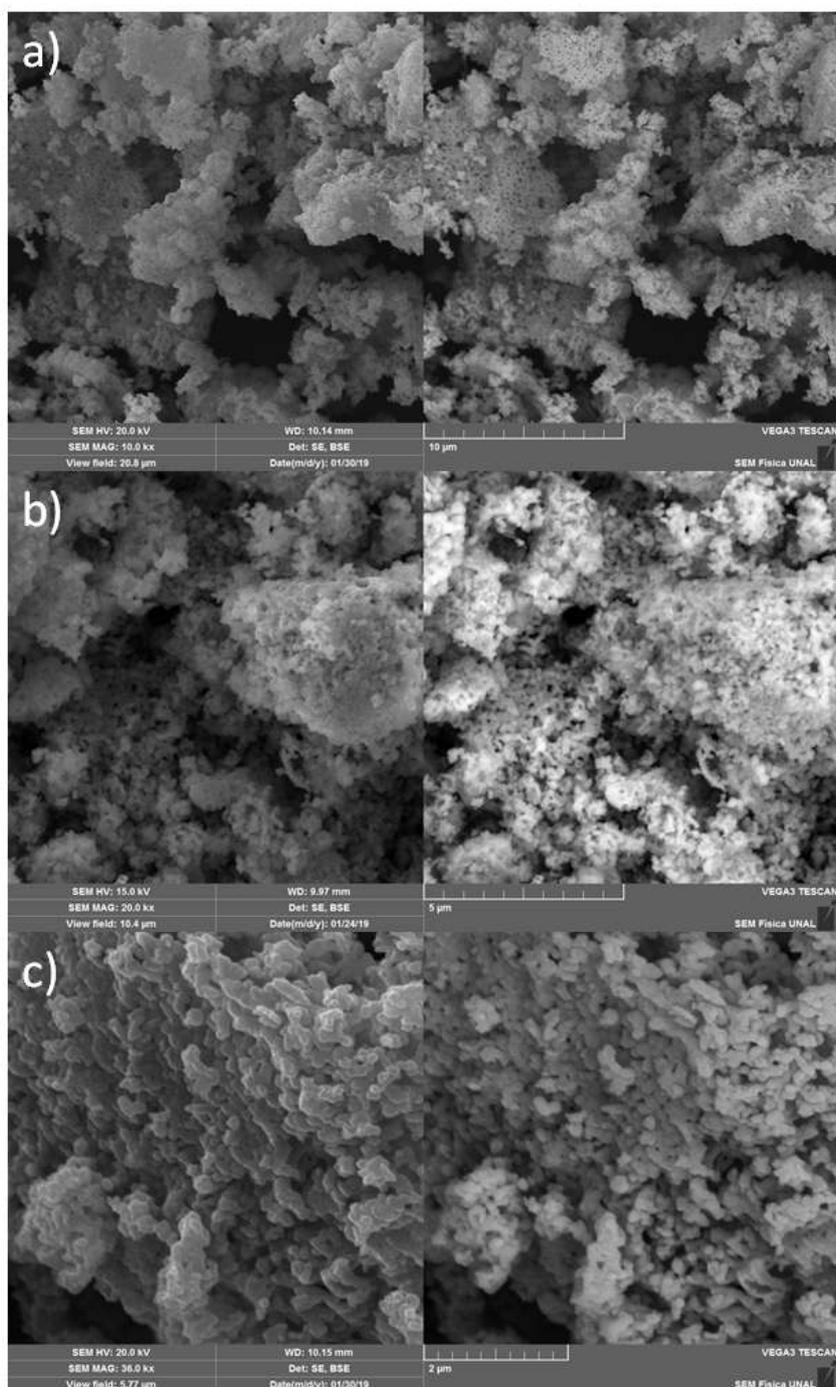

Fig. 4. Surface SEM images for the $Sr_2FeWO_6$ samples. Left pictures correspond to secondary electron images and right to backscattering images with magnifications (a) 10kX, (b) 20kX and (c) 36kX.

The maximum magnification analyzed in figure 4c (36 kx) allows us to conclude that the grains are definitely nanometric in size, forming agglomerates of ciliary shape whose length reaches between *1* and *4 μm*, formed by grains strongly diffused among themselves, but with a marked tendency porous between agglomerates, such that the pore sizes can exceed the lengths of these agglomerates. This distribution has to do with the synthesis technique, since in the combustion process nanoparticles are formed that are collected in the form of ash that does not have a compact volumetric structure, which gives rise to the spongy characteristic, due to the growth of the crystallites and particles in an almost independent manner that ends up being partially densified in the final thermal sintering process, which takes place for a short time and without the application of hydrostatic pressure on the sample. As showed in figure 5, the sizes of the particles, observed in the SEM images, were established experimentally through the ImageJ software [37], obtaining a mean size value of *160±5 nm*, which suggests that the combustion is a good synthesis technique for the production of nanometric materials of this perovskite family.

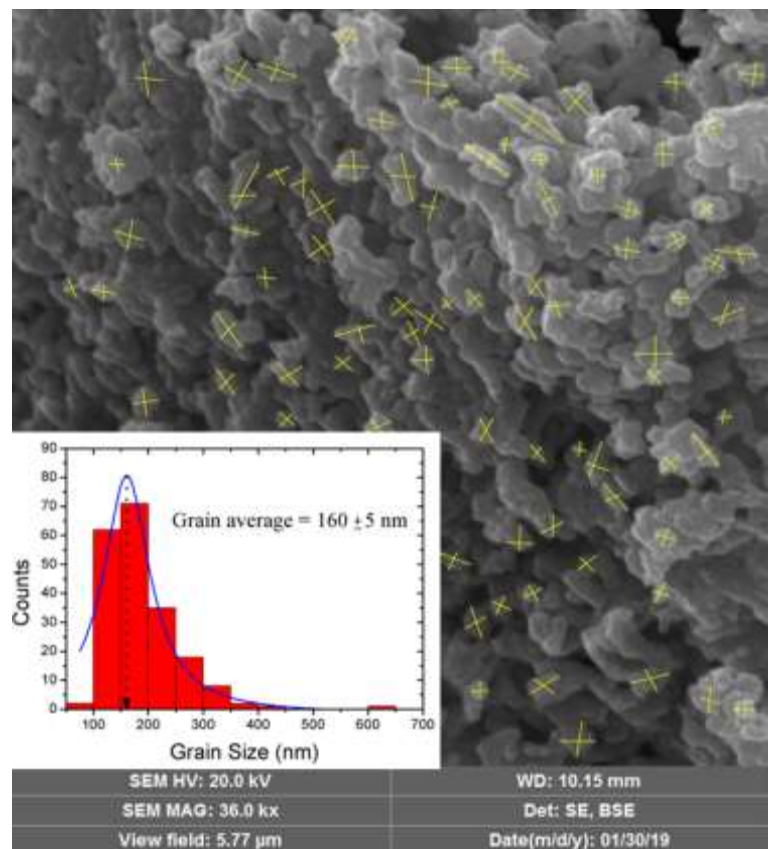

Fig. 5. Mean grain size obtained from the SEM images.

Finally, in relation to the analyzed morphology it is possible to affirm that practically all the surface shows the same distribution, for which it is very difficult to observe any evidence of the presence of the ternary compound $SrWO_4$, because its proportion in the material is negligible and only it was established because of the excellent quality of the structural analysis.

Concerning the semiquantitative study of the composition of the material, the EDX analyzes exemplified through the spectrum of figure 6 reveal that, at least superficially, there are no other elements other than those expected from the precursor compounds. In the inset of Figure 6 a table is presented in which the degree of coincidence, between the percentage weights expected from the stoichiometric calculation on the formula $Sr_2FeWO_6$ and the estimated percentage weights from the deconvolution of the energy spectrum for the material, evidences a medium matching of 90%. This degree of coincidence is good, considering that the calculation was made from the formula of the double perovskite, without considering the deviation due to the presence of the ternary phase $SrWO_4$ or that due to the weight light character of the oxygen.

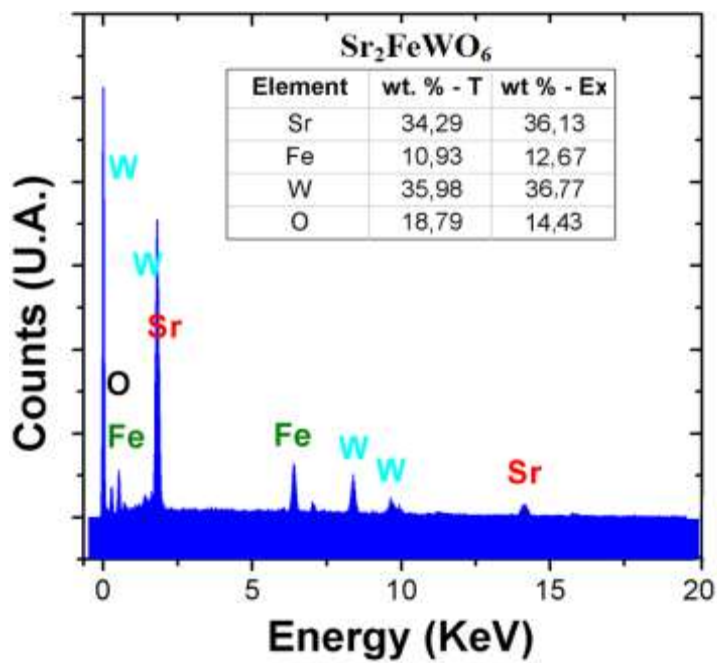

Fig. 6. EDX spectrum and weight percentages of Sr, Fe, W and O in the material.

*C. XPS and Mössbauer spectroscopy characterizations*

Because one of the objectives of this document is to establish the oxidation states of the Fe and W cations, XPS spectroscopy analysis was performed. The spectrum obtained for the $Sr_2FeWO_6$ material in the energy range between *0* and *1200 eV* is shown in figure 7. The spectrum reveals only the presence of the elements that make up the material, which is consistent with the results obtained by the EDX technique, where the presence of impurities was not observed. However, a deficiency in the Fe and W contents on the surface of the material can be evidenced, which can be attributed to the segregation of Sr atoms that takes place during the synthesis process and which has also been reported by other authors who have applied conventional synthesis methods [21].

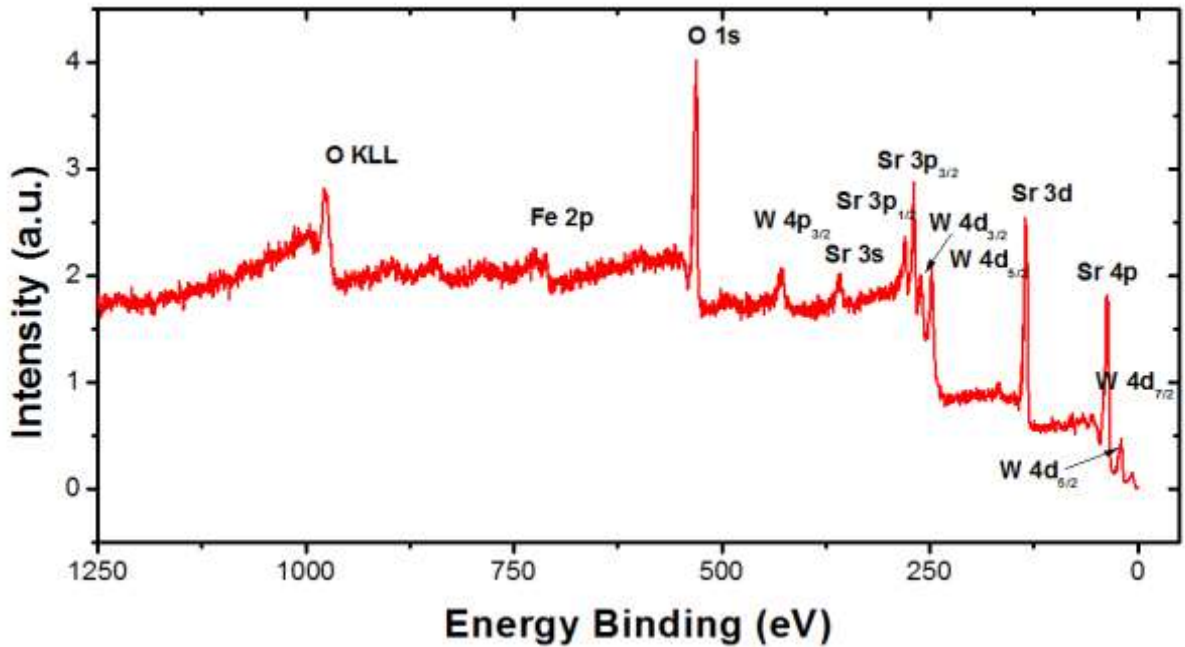

Fig. 7. XPS spectrum for the $Sr_2FeWO_6$ double perovskite.

Since the XPS technique is sensitive on the surface, the depth and the average free path of the leaking electron, which corresponds approximately to a length between *10* and *100 Å*, limit this sensitivity. A rigorous treatment of XPS spectra in transition metal oxides must take into account the effects of charge transfer of many bodies, which are primarily responsible for presenting a characteristic division of the 2p spectra of the compounds [38]. Additionally, a semiquantitative analysis of the valences of the ions present on the surface can be performed. In figure 8, the XPS spectra are shown separately for W-4f, Fe-2p and O-1s. From the deconvolution adjustment made by means Gaussian and Lorentzian functions for each of these spectra is possible to determine the presence of ions with different valences, both for W cations and Fe cations. In the case of W ions (figure 8a), the spectrum shows two doublets that related to spin-orbit splitting for W-4$f_{7/2}$ and W-4$f_{5/2}$, corresponding to different valences.

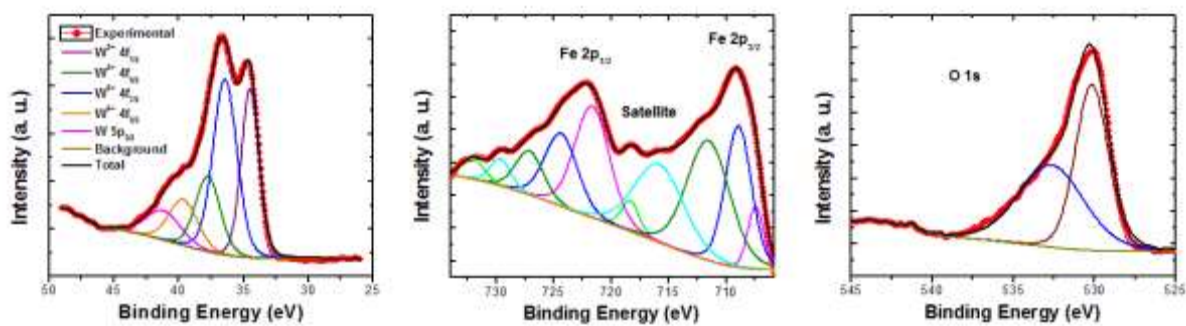

Fig. 8. Separate XPS spectra for the $Sr_2FeWO_6$ double perovskite: a) W-4f, b) Fe-2p and c) O-1s.

These peaks have been studied and characterized considering the different valence values for the case of W ions. For example, Moulder et al. report that for $W^{6+}$ ions the peaks corresponding to W-4$f_{7/2}$ and W-4$f_{5/2}$ are observed for bond energy values *35.6 eV* and *37.7 eV*, respectively [39]. On the other hand, for $W^{5+}$ ions these peaks are reported for slightly lower energies, such as those reported by Zhang et al. at *34.9 eV* and *37.1 eV*, respectively [40]. In the case of our W spectra, the presence of two doublets resulting from spin-orbit splitting, corresponding to ions with valences $W^{5+}$ and $W^{6+}$, as well as the presence of a peak corresponding to W-5$p_{3/2}$ can be evidenced. The peaks observed in *34.5 eV* and *37.7 eV* correspond to $W^{5+}$ ions, while the double peak located in *36.4 eV* and *39.6 eV* corresponds to $W^{6+}$ [41]. This combination of ions not yet reported for this material can be attributed to the thermodynamic effects that occur at the time of combustion. On the other hand, the appearance of $W^{6+}$ and $W^{5+}$ ions induces the presence of $Fe^{2+}$ and $Fe^{3+}$ ions in the structure so that the ionic neutrality of charges is maintained. This fact is seen in figure 9, where the spectrum corresponding to the Fe-$2p_{1/2}$ and Fe-$2p_{3/2}$ peaks presents contributions for both electronic valences. For this deconvolution process the mechanism proposed by Grosvenor was followed [42], finding a majority contribution for $Fe^{2+}$, close to 66.9%, in contrast to 33.1% for $Fe^{3+}$. The spectrum observed in Figure 8b has characteristics that have been reported for $Fe^{2+}Cr_2S_4$ [43] and $Fe^{3+}_2O_3$ [44], both for the characteristic peaks of Fe-$2p_{1/2}$ and Fe-$2p_{3/2}$, and for the additional peaks, where the link energy values for these peaks, *715.8 eV* and *718.3 eV*, are found for intermediate values of those reported for the reference spectra of $Fe^{2+}$ and $Fe^{3+}$ [45-46]. The characteristic spectrum for O-1s observed in Figure 8c shows the presence of oxygen ions in two different states, with binding energies of *530.1 eV* and *532.6 eV*. The peak of lower binding energy corresponds to oxygen in the perovskite structure [39, 47], while the second peak, seen as a slight shoulder over the previous one, is related to the occurrence of an eventual OH group on the surface of the sample [48] that would be produced by the method of synthesis.

On the other hand, it is known that in a spectrum obtained by the Mössbauer technique, three parameters that demonstrate the characteristics of the material under study can be determined: isomeric displacement (IS), quadrupole splitting (QS) and hyper magnetic displacement (Hhf). The isomeric displacement provides information on the electronic density, and therefore, on the valence and coordination number. Quadrapolar splitting is a measure of how the electric field gradient around the atom affects the nuclear energy levels of the atom, and like isomeric displacement, provides information on the oxidation state. Finally, hyper fine magnetic displacement is a measure of the magnetic interaction between the nuclear magnetic moment and the magnetic field external to the nucleus, also known as the

Zeeman nuclear interaction [49]. Figure 10a depicts the Mössbauer spectrum of $Sr_2FeWO_6$ at a temperature of 3 K. The spectrum is modeled with a sextuplet, which is evidence of magnetically ordered domains at this temperature. According to the digital printing characterization method [28], which consists in comparing the reported values for isomeric displacement, quadrupole splitting and hyperfine magnetic displacement with the values reported for standard materials, it is possible to determine the valence of Fe ions present in the material.

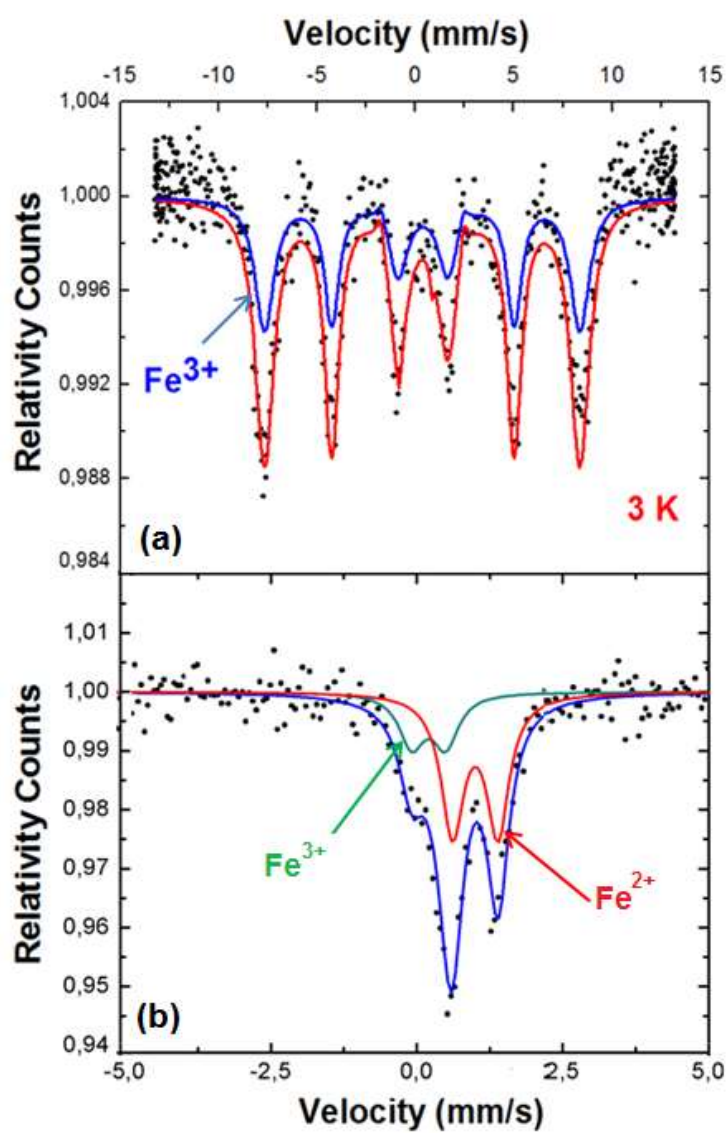

Fig. 9. Mössbauer spectra for the Sr2FeWO6 material performed at low temperature (a) and at room temperature (b).

The values of isomeric displacement (0.36 mm/s), quadrupole sppliting (*-0.04 mm/s*) and hyperfine magnetic displacement (*497.5 KOe*) in the sextuplet are evidence of the presence of $Fe^{3+}$ ions that are the cause of the properties magnetic material at this temperature [50]. Meanwhile, in figure 10b the Mössbauer spectrum for $Sr_2FeWO_6$ is presented at room temperature, which is constituted by two doublets as an evidence of the presence of $Fe^{2+}$ and $Fe^{3+}$ ions in the material. This fact is corroborated by the values obtained for the parameters of the isomeric displacement and the quadrupole unfolding for each of the doublets, in accordance with the results previously obtained by XPS. The spectrum of greater relative area (68%) corresponds to the presence of $Fe^{2+}$ ions, with IS = 1.0 and QS = 0.77, while the spectrum of lower relative area (32%) corresponds to the presence of $Fe^{3+}$ ions, with IS = 0.20 and QS = 0.57 [50].

*D. Magnetic response and structure*

In order to evaluate the magnetic response of the material, measurements of magnetization AC as a function of temperature under the application of magnetic field intensities *H = 10 Oe* in a frequency of *1000 Hz*. Other authors have reported the magnetic response of $Sr_2FeWO_6$ materials synthesized by conventional methods, such as solid-state reaction, exhibiting antiferromagnetic responses with Néel temperatures close to *35 K* [20, 51]. Additionally, the analysis of the magnetic phase transitions that occur in this type of materials based on stoichiometric substitutions with transition metal cations, for example materials of type $Sr_2FeMo_{1-x}W_xO_6$, have revealed the occurrence of a coexistence of antiferromagnetic and ferrimagnetic features, as a result of the variation in the valences of the Fe and W ions [52]. On the other hand, comparative studies of the magnetic response in this type of materials have shown that the sensitivity of this property can change depending on the thermal procedure and the atmospheres used in the sintering processes [22]. The behavior of the magnetic response of the $Sr_2FeWO_6$ material is presented in figure 10a. At low temperatures the magnetic response of the material exhibits an increasing behavior, as the temperature increases, reaching a maximum value for temperatures close to 100 K, while for higher temperatures its behavior is decreasing, resembling a characteristic behavior of a paramagnetic material. The behavior of the derivative of susceptibility as a function of temperature is presented in Figure 10b, where a maximum close to *T=32 K* is observed. This value is consistent with others previously reported for the Néel temperature in these materials, where the antiferromagnetic interaction is due to $Fe^{2+}$ cations ($3d^6$, S=2, high spin configuration) with $W^{6+}$ ions ($4f^{14}$, S=0), through double exchange mechanisms, which are mediated by $O^{2-}$ anions.

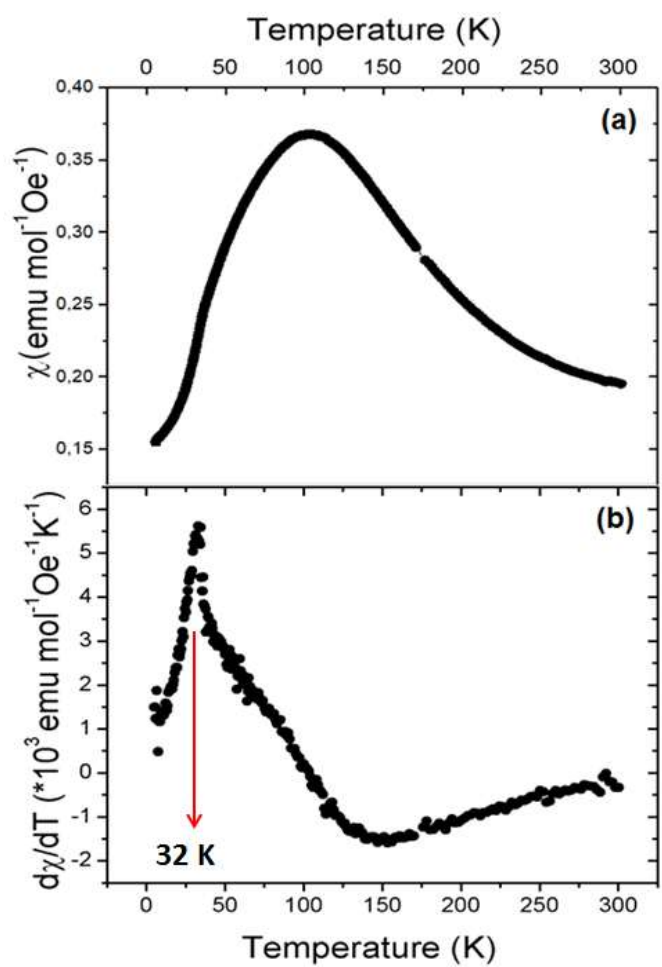

Fig. 10. Real AC magnetic susceptibility as a function of temperature measured for the $Sr_2FeWO_6$ material (a) and temperature derivative of magnetic susceptibility (b).

Additionally, a minimum value is presented for temperatures close to *150 K*, which could exhibit the presence of a ferromagnetic phase, which is maintained for higher temperatures. For values close to room temperature, the ferromagnetic response is maintained, so that the material does not reach the paramagnetic state at that temperature. These results are in agreement with the results obtained from XPS and EDX, where the presence of $Fe^{3+}$ ($3d^5$, S=5/2 high spin configuration) and $W^{5+}$ ($4f^{13}$, S=1/2) ions was evidenced at room temperature, which they are the cause of producing the ferromagnetic response in the material.

## IV. Conclusions.

The assisted gel combustion synthesis method emerges as a very valuable alternative in the production of perovskite-type materials that need reducing atmospheres to work by means of conventional synthesis methods, offering advantages as energy saving and decreasing in the preparation time. The Rietveld refinement performed from the XRD pattern determined that the $Sr_2FeWO_6$ material crystallizes in a primitive monoclinic structure, $P2_1/n$ space group, with evidences of cationic ordering with respect to the Fe and W ions. The distortions that occur in the octahedral substructures $Fe-O_6$ and $W-O_6$ are more notable in the direction [001]. By means of scanning electron microscopy images an average surface grain size of nanometric order (160±5 nm) was established, with intergranular spacing attributed to the low volumetric density of the material. The X-ray energy dispersion analysis by electrons showed that the samples evaluated do not show impurities of other chemical elements outside those expected from the stoichiometric formula. The X-ray photoelectron spectroscopy analysis for Fe and W ions showed the presence of different valences for each of these elements, determining a majority concentration of $Fe^{2+}$ ions (66.9%) in contrast to $Fe^{3+}$ ions (33.1%), which was corroborated by analyzing the spectra obtained by Mössbauer spectroscopy.

Likewise, an antiferromagnetic behavior was established at low temperatures (with $T_N=32\ K$), which is attributed to the presence of $Fe^{2+}$ ions, with a ferromagnetic tendency at high temperatures (above $150\ K$) due to the $Fe^{3+}$ ions. These results suggest that two crystalline phases coexist in the material: 66.9% of $Sr_2^{2+}Fe^{2+}W^{6+}O_6^{2-}$, whose antiferromagnetic character is dominant at low temperatures, and 33.1% of $Sr_2^{2+}Fe^{3+}W^{5+}O_6^{2-}$, whose ferromagnetic nature is more relevant at high temperatures.

## V. Acknowledgments.

This work was partially supported by Division of Investigation and Extension (DIEB) of the National University of Colombia and Departamento Administrativo de Ciencia y Tecnología Francisco José de Caldas – COLCIENCIAS, on the project FP80740-243-2019. One of us (J.I. Villa Hernández) received support by Departamento Administrativo de Ciencia y Tecnología "Francisco José de Caldas", COLCIENCIAS, on the scholarship program for national doctorates. In the same way, J.I. Villa-Hernández would like to acknowledge S.C. Tidrow for partial financial support through the Inamori Professorship funds of S.C. Tidrow.